# Complex dynamics in generalizations of the Chaplygin sleigh

## S. P. Kuznetsov

*Udmurt State University, Izhevsk, Russia*
spkuz@yandex.ru

The article considers Chaplygin sleigh on a plane in potential well, assuming that an external potential force is supplied at the mass center. Two particular cases are studied in some detail, namely, a one-dimensional potential valley and a potential with rotational symmetry; in both cases the models reduce to four-dimensional differential equations conserving mechanical energy. Assuming the potential functions quadratic, various behaviors are observed numerically depending of the energy, from those characteristic to conservative dynamics (regularity islands and chaotic sea) to strange attractors. This is another example of nonholonomic system manifesting these phenomena (similar to those for Celtic stone or Chaplygin top), which reflects a fundamental nature of these systems occupying intermediate position between conservative and dissipative dynamics.

## 1. Introduction

Study of complex dynamics of nonlinear systems, including dynamic chaos, is a fundamental interdisciplinary problem.

In mechanics, besides the systems described in the framework of the Hamiltonian formalism (conservative) and systems with friction (dissipative), a class of systems with nonholonomic constraints is of special interest [1,2]. They include many situations of great practical value, for example, in the analysis of mobile vehicles, e.g. in the context of robotics. Hierarchy of nonholonomic systems include a variety ranging from simple (integrable) to complex (nonintegrable) cases [3]. A representative example of the complex dynamics is motion of a solid body with a convex smooth surface on a rough plane (the rattleback, or the Celtic stone). Its fundamental property, like in other nonholonomic systems occupying a similar place in the hierarchy, is lack of invariant measure in the sense of the Liouville theorem [4]. Although the system is conservative (conservation of mechanical energy) and symmetric with respect to time reversal, the phase volumes during the dynamic evolution do not remain constant, undergoing locally compression or expansion in the phase space. Due to this, the asymptotic behaviors associated with attractors can occur, like those in dissipative systems [5-7].

One of the simplest paradigmatic examples in the nonholonomic mechanics is Chaplygin sleigh that is a platform that can move on a plane surface, having a "knife edge" attached to the sleigh as one of the supports, capable of sliding only in the longitudinal direction. The dynamics of the classical model with some initial translational and angular velocity leads to arising steady motion of the sleigh at a constant speed along the direction of the knife edge that corresponds to a simple attractor of the dynamical system.

The present paper examines modifications of the Chaplygin sleigh problem, which make possible complex dynamics with conservation of mechanical energy. Namely, we consider Chaplygin sleigh on a plane in a potential field that ensures restriction of the motions in one or two dimensions assuming that the potential force is supplied at the mass center. The complex dynamics can arise due to the fact that when sliding down in the potential field and then moving up by inertia, the sleigh tends to orient the knife edge to be back respective to the mass center. After the body begins to slide in opposite direction, it tends to make a turn to have the knife edge back again. Nevertheless, with a relatively small energy the resulting motions appear to be quasi-periodic, but at sufficiently large energies the chaotic motions become typical.

## 2. Basic equations

Consider Chaplygin sleigh on a plane (Fig. 1) using a laboratory frame ($x$, $y$) and a frame ($X$, $Y$) fixed on the platform. The condition of the nonholonomic constraint is that the velocity direction for the point $A$ is fixed relatively the sleigh. This can be interpreted as a knife-edge attached at $A$, which is allowed to slide along, while transversal motions are prohibited. We assume that the knife-edge direction is the axis $X$, and the center of mass $C$ is located on the same axis at a distance $a$. The reaction force, which prevents transverse motions of the knife-edge, is directed along the $Y$ axis.

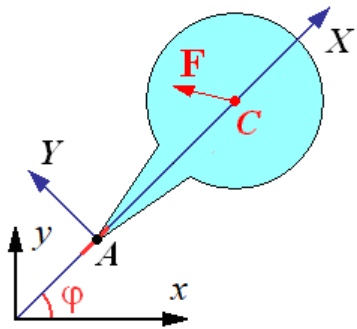

Fig.1. Chaplygin sleigh on the plane under action of potential force **F**. *C* is the mass center, and *A* is a point of location of the nonholonomic constraint allowing motion of this point exclusively along the direction of the knife-edge shown as a red segment.

The equations of motion are

$$m\dot{u} = ma\omega^2 + F_x \cos\varphi + F_y \sin\varphi,$$
$$(J + ma^2)\dot{\omega} = -ma\omega u + a(-F_x \sin\varphi + F_y \cos\varphi),$$
$$\dot{\varphi} = \omega, \qquad (2.1)$$
$$\dot{x} = u\cos\varphi - a\omega\sin\varphi,$$
$$\dot{y} = u\sin\varphi + a\omega\cos\varphi.$$

Here $m$ is mass of the sleigh, $J$ is moment of inertia, $\varphi$ is rotation angle of the sleigh, $u$ is velocity of the knife-edge, $\omega$ is angular velocity of the sleigh, $x$ and $y$ are coordinates of the mass center in the laboratory frame. The external force components are expressed as derivatives of the given potential function: $(F_x, F_y) = (-\partial_x U(x,y), -\partial_y U(x,y))$. The system has an integral of motion expressing the conservation of mechanical energy:

$$\tfrac{1}{2}mu^2 + \tfrac{1}{2}(J+ma^2)\omega^2 + U(x,y) = \text{const} \qquad (2.2)$$

and is invariant with respect to the inversion of time, the involution

$$t \to -t,\ u \to -u,\ \omega \to -\omega. \qquad (2.3)$$

We will examine two specific problems that can be reduced to smaller number of equations. The first relates to the case when the potential function depends only on one coordinate in the laboratory reference frame, and the second to the situation when the potential function has rotational symmetry. For simplicity and concreteness, we limit consideration with the potentials given by quadratic functions. If one excludes a nonholonomic constraint, then the problems reduce to a one-dimensional or a two-dimensional linear oscillator, respectively.

Let the potential be given first by the expression $U = \tfrac{1}{2}ky^2$, where $k$ is a constant. Substituting in (2.1) $F_x = 0$, $F_y = -ky$, using dimensionless variables and parameters

$$t = t'\sqrt{k/m},\ u' = ua^{-1}\sqrt{m/k},\ \omega' = \omega\sqrt{m/k},\ y' = y/a,\ x' = x/a,\ \mu = 1 + J/ma^2, \qquad (2.4)$$

and omitting primes for brevity we arrive at the closed set of four equations

$$\dot{u} = \omega^2 - y\sin\varphi,$$
$$\mu\dot{\omega} = -\omega u - y\cos\varphi,$$
$$\dot{\varphi} = \omega, \qquad (2.5)$$
$$\dot{y} = u\sin\varphi + \omega\cos\varphi.$$

The motion along the $x$ axis is determined by an additional separate relation

$$\dot{x} = u\cos\varphi - \omega\sin\varphi. \qquad (2.6)$$

The integral of motion expressing conservation of the dimensionless energy *W* reads

$$W = \tfrac{1}{2}(u^2 + \mu\omega^2 + y^2) = \text{const} . \tag{2.7}$$

Consider now a potential with rotational symmetry setting $U = \tfrac{1}{2}k(x^2 + y^2)$ and substituting $F_x = -kx$, $F_y = -ky$ in (2.1). In dimensionless quantities (2.4), after simple transformations with the change of variables

$$x = \xi\cos\varphi - \eta\sin\varphi, \quad y = \xi\sin\varphi + \eta\cos\varphi \tag{2.8}$$

we come to a system of four equations, where the angular coordinate φ is excluded:

$$\begin{aligned}&\dot{u} = \omega^2 - \xi, \quad \mu\dot{\omega} = -\omega u - \eta,\\ &\dot{\xi} = u + \omega\eta, \quad \dot{\eta} = a\omega - \omega\xi.\end{aligned} \tag{2.9}$$

In this case, the energy integral has a form

$$W = \tfrac{1}{2}(u^2 + \mu\omega^2 + \xi^2 + \eta^2) = \text{const} . \tag{2.10}$$

## 3. Dynamics in the case of potential depending on one coordinate

Let us study movement of the Chaplygin sleigh in the case of quadratic potential depending on one coordinate by means of numerical simulation by the fourth-order Runge – Kutta method. The main parameter, on which the character of the dynamics depends, is the energy *W*, the value of which is determined by setting initial conditions for the variables *u*, ω, *y* (see (2.7)). The parameter μ is assumed to be fixed, namely, μ=10.

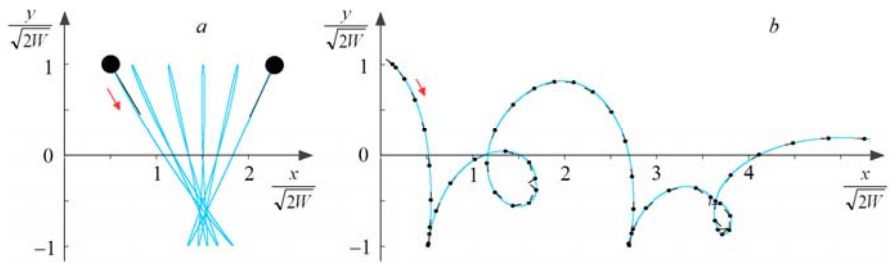

Fig.2. Regular motion of the Chaplygin sleigh in laboratory frame at low energy *W*=0.0625 (*a*) and complex motion at high energy *W*=62.5 (*b*) in the case of potential depending on one coordinate.

Just preliminary calculations of trajectories by integrating equations (2.5) and (2.6) show that for low energies typical are regular motions, and for high energies chaotic ones are characteristic (Fig. 2). In the first case, the sleigh performs oscillations in the potential profile of such kind that it does not have enough time to turn around during the characteristic period, and the movements alternate in direction when the center of mass is ahead of the knife-edge and vice versa. In the second case, the instability of the motion of the knife-edge forward develops to a significant extent, which leads to pronounced rotations of the sleigh and to chaotic dynamical behavior.

Actually, the regular or chaotic nature of the motion is determined by the four-dimensional autonomous dynamical system (2.5) on the three-dimensional manifold of constant energy. Figure 3 shows diagrams that are obtained in a two-dimensional section of the phase space corresponding to the moments of passage of the variable *y* through zero value (in the direction from positive to negative).

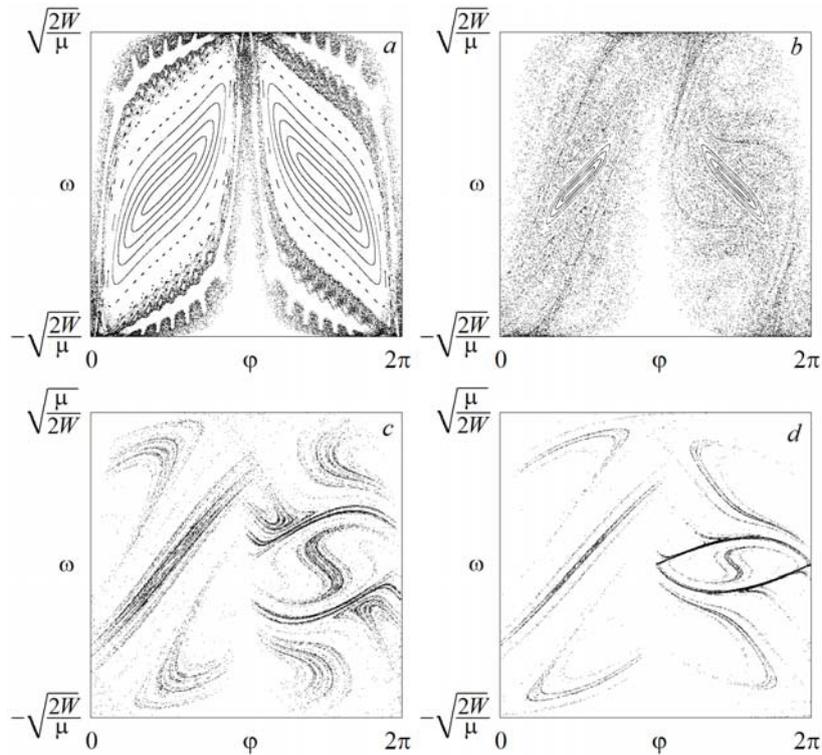

Fig.3. Phase portraits of model (2.5) in cross-section $y = 0$ (only points where $\dot{y} < 0$ are shown). Parameters: $\mu = 10$, $W$=0.0125 (*a*), 1.25 (*b*), 25 (*c*), 62.5 (*d*).

At small energies *W*, regular quasiperiodic motions dominate, which are represented by closed invariant curves, and chaos occurs in narrow regions (stochastic layers) separating the regions of regular dynamics. With the growth of the energy, the domains of chaotic motions enlarge, forming a "chaotic sea" surrounding the surviving islands of regularity. Although the dynamics look similar to those in systems conserving phase volume [8,9], the chaotic sea here appears to represent not a set invariant under time reversal, but rather has to be treated as a kind of attractor ("fat attractor") [10]; the distinction is that the positive and negative Lyapunov exponents are not equal in absolute value (see Table 1). With growth of the energy, this difference becomes more pronounced, and the sustained chaos is determined by actual strange attractors similar to those in dissipative chaotic systems [11, 12]; they have subtle filament transversal structure and are characterized by fractal dimensions, which is between 1 and 2 for the cross-section representation of them.

Concerning the Lyapunov exponents of the system (2.5) one should note immediately that for any sustained motion there must present two zero exponents, one is associated with a perturbation along the phase trajectory, as usual in autonomous systems, and other is associated with an energy shift in the system with energy conservation. For the complete system (2.5), (2.6), one more zero Lyapunov exponent takes place due to the translational invariance along the *x* axis.

Calculation of the Lyapunov exponents with traditional numerical method [13] for closed invariant curves shows that four exponents are zero (up to computational errors). For chaotic motions nonzero Lyapunov exponents are given in Table 1. Presence of a positive exponent indicates chaotic nature of the dynamics. Observe that sums of the positive and negative exponents are negative that means that the chaotic sets have to be interpreted as attractors. In the last row in the table the Kaplan – Yorke dimensions [11, 12] of the attractors in the cross-sections (Fig. 3), which are expressed in this case as $D_{KY}=1+\lambda_+/|\lambda_-|$; for fat attractor (the first column) it is close to 2.

Table 1. Nonzero Lyapunov exponents of chaotic motions in model (2.5) at μ=10

| W | 1.25 | 25 | 62.5 |
|---|---|---|---|
| λ | 0.1093 −0.1131 | 0.0508 −0.0806 | 0.0399 −0.1009 |
| $D_{KY}$ | 1.97 | 1.63 | 1.39 |

An interesting feature of the system under consideration is that chaotic dynamics of the reduced system (2.5) give rise to a one-dimensional random walk of diffusion type [14] for the sleigh in the laboratory frame along the axis *x*. It is illustrated in Fig. 7. Panel (*a*) shows how the sleigh trajectories look like. The diagram is plotted basing on numerical integration of Eqs. (2.5), (2.6) for the launch at *x*=0 with arbitrary remaining initial conditions, tracking the motion up to *t*=80; further, the computations continue with the current *u*, ω, φ and *y*, but with a launch again from *x*=0. Thus, it is a family of fragments of one and the same trajectory in the sense of the reduced equations corresponding to an orbit tending to chaotic attractor of Fig. 3*d*. As one can judge, the observed motion is a random walk, where the distribution of distances from start to finish reached at a fixed time interval *t* tends asymptotically to the Gauss distribution with variance $\sigma^2=\tfrac{1}{2}Dt$, where *D* is the diffusion constant. Panel *b* shows the variance dependence on *t* and panel *c* shows the distribution function for particular $t=50$ as obtained by data processing for a large ensemble of samples. From panel *b* one can see that the dependence in the double logarithmic scale is well fitted by a straight line $\lg\sigma^2 = \lg t + 2.360$ with unit slope, and the shift parameter allows estimating the diffusion coefficient: $D\approx 114$. In panel *c* one can observe a very good correspondence of the empirical distribution (shown in gray) and the Gaussian one (blue curve) with the variance obtained from the data processing.

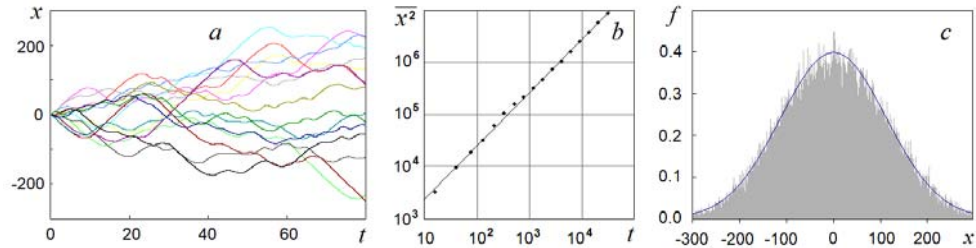

Fig.4. A set of trajectories illustrating random walk of the sleigh along the potential valley (*a*), dependence of the variance of the distance on time in double logarithmic scale (*b*) and the distribution function obtained from the numerical data processing at particular $t=50$ in comparison with the Gaussian distribution $f = (2\pi)^{-\frac{1}{2}}\exp(-x^2/2\sigma^2)$ at $\sigma=112.8$ (*c*).

## 4. Dynamics in potential with rotational symmetry

Let us turn now to results of numerical study of the Chaplygin sleigh motion in potential field with rotational symmetry.

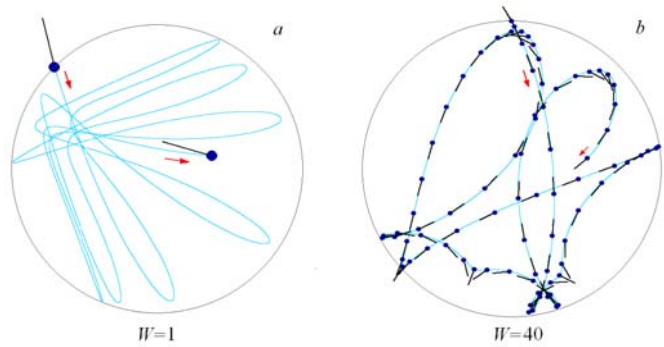

Fig.5. An example of Chaplygin's sleigh regular motion in potential well with rotational symmetry typical for low energy (*a*), and an example of complex motion at relatively high energy (*b*) on the plane (*x*, *y*). Radius of the disk, in which the mass center moves, is $\sqrt{2W}$.

Figure 5 shows typical pictures of movements in a laboratory frame observed at low and high energy, when the motion is regular or chaotic, respectively. Trajectories are obtained by numerical integration of equations (2.9) with recalculation of ξ and η to the coordinates in the laboratory frame *x* and *y* by means of (2.8).

Figure 6 shows diagrams illustrating the dynamics for different energies on the coordinate plane ξ, η in cross-section of the phase space by a zero-velocity hypersurface $u=0$. At low energy, one can see a set of invariant curves corresponding to quasiperiodic dynamics. There are fixed points of elliptic type surrounded by the invariant curves. With increasing energy, areas of chaos appear as stochastic layers that expand to form the chaotic sea surrounding the surviving islands of regular dynamics. Regular movements correspond to trajectories with four zero Lyapunov exponents, and chaotic ones correspond to trajectories that have one positive and one negative Lyapunov exponent equal in absolute values (at least within the limits of the calculation error). This picture corresponds qualitatively to that observed in Hamiltonian systems [8, 9]. As the energy increases, differences become noticeable. In particular, the positive and negative exponents for chaotic movements become different from each other in absolute value, so the set of chaotic trajectories should be interpreted as an attractor, more precisely, as a "fat attractor", because the dimension in the cross-section is close to 2. With even greater energy, chaos is observed, corresponding to usual strange attractors, similar to those observed in dissipative systems, which are characterized by filamentary fractal structure. Table 2 summarizes the non-zero Lyapunov exponents observed at relatively high energies, and the estimates of fractal dimension of the attractor in the Poincaré section according to Kaplan – Yorke [11,12].

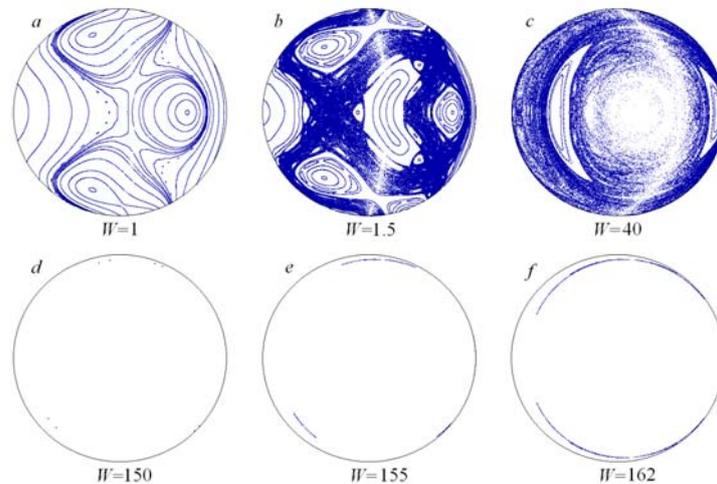

Fig.6. Phase portraits of the Chaplygin sleigh model in a potential field with rotational symmetry (2.9) on a plane of variables (ξ, η) in cross-section $u = 0$ at various values of energy *W*. Regular dynamics (*a*), chaotic sea and regularity islands like those observed in conservative systems with complex dynamics (*b*, *c*), an attractor in the form of periodic point (*d*) and strange attractors born according to the Feigenbaum scenario (*e, f*) are observed. The axes of the frame ξ, η are parallel to the axes of the moving reference frame *X, Y*, but the origin is placed at the origin of the laboratory frame *x, y*.

Interestingly, in the region $W = 120 \div 160$ in the system a transition to chaos is observed through the Feigenbaum period doubling bifurcation cascade [11, 12]. The estimated convergence constant to the accumulation point corresponds to Feigenbaum's number $\delta = 4.69...$. Such a transition is well known and typical in dissipative nonlinear systems, but for the model with energy conservation and time reversal symmetry it deserves to be specifically emphasized. (Like that noted earlier for the Celtic stone model [6, 7].) In Fig. 7, the period doubling cascade is illustrated with a traditional picture of bifurcation tree. The energy parameter is plotted along the horizontal axis, and values of one of the dynamical variables are plotted along the vertical axis corresponding to moments of passages of the cross-section *u*=0 in the sustained dynamical regime. A period-doubling bifurcation looks like a split of the branches of the "tree", and the chaotic modes correspond to the dotted areas of the "crown".

Table 2. Nonzero Lyapunov exponents of motions in model (2.5) at μ=10

| $W$ | 1.5 | 40 | 150 | 155 | 162 |
|---|---|---|---|---|---|
| $\lambda$ | 0.0095 | 0.1616 | –0.1015 | 0.0542 | 0.1147 |
|  | –0.0095 | –0.2041 | –0.5612 | –0.7332 | –0.8230 |
| $D_{KY}$ | 2.0 | 1.79 | 0 | 1,07 | 1,14 |

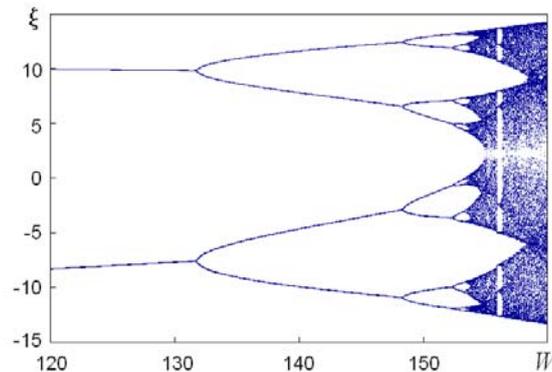

Fig.7. Diagram illustrating transition to chaos through period doubling bifurcations depending on the energy for the Chaplygin sleigh in potential with rotational symmetry (2.9), at $\mu = 10$.

## 5. Conclusion

Two models based on the Chaplygin sleigh are considered providing interesting examples of four-dimensional systems characterized by presence of an energy integral and reversibility in time. These examples clearly demonstrate the fundamental feature of nonholonomic systems, namely, their intermediate position between conservative and dissipative systems with combination of phenomena of nonlinear dynamics inherent to them. Similar phenomenology was observed previously for the Celtic stone and Chaplygin's top [5-7], as well as for some non-autonomous models [10, 15], but the systems considered here look simple and natural, and they are autonomous. The results obtained are interesting in concern of general methodology of the nonholonomic mechanics, as well as for possible applications for creating and control of mobile devices, such as wheeled vehicles basing on the principles of nonholonomic mechanics.

# Funding

The work was supported by Russian Science Foundation, grant 15-12-20035.